\documentclass[a4paper,11pt]{article}
\usepackage{pos}
\usepackage{physics}
\usepackage{dsfont}
\usepackage{amsmath}
\usepackage{amsfonts}
\usepackage{bigints}
\usepackage{booktabs}

\title{Noise-Aware Mixed-State Quantum Computation via Parameterized Quantum Channels}

\author*[a]{Giuseppe Clemente}
\author[a]{Kevin Zambello}

\affiliation[a]{Dipartimento di Fisica dell'Universit\`a di Pisa and INFN --- Sezione di Pisa, Largo Pontecorvo 3, I-56127 Pisa, Italy.}

\emailAdd{giuseppe.clemente@unipi.it}
\emailAdd{kevin.zambello@unipi.it}

\abstract{
    Non-unitary protocols are already at the base of many hybrid quantum computing applications, especially  in the noisy intermediate-scale quantum (NISQ) era where quantum errors typically affect the unitary evolution. However, while the framework for Parameterized Quantum Circuits is widely developed, especially for applications where the parameters are optimized towards a set goal, we find there are still interesting opportunities in defining a unified framework also for non-unitary protocols in the form of Parameterized Quantum Channels as a computing resource. We first discuss the general parameterization strategies for controlling quantum channels and their practical realizations. 
Then we describe a simple example of application in the context of error mitigation,
where the control parameters for the quantum channels are optimized in the presence of noise, 
in order to maximize channel fidelity with respect to a given target channel.}

\FullConference{The 41st International Symposium on Lattice Field Theory (LATTICE2024)\\
 28 July - 3 August 2024\\ Liverpool, UK\\}

\begin{document}
\maketitle

\section{Introduction}\label{sec:intro}

Despite continuous progress, achieving quantum advantage in the NISQ era remains a daunting task due to the limitations of protocols based on ideally noiseless unitary circuits. 
In general, real quantum hardware introduces noise, turning ideal unitary operations into non-unitary quantum channels. This makes it useful to rethink any ideal computation protocol with a realistic one, where parametric unitary circuits $U$ are replaced by parametric quantum channels $\mathcal{E}$ to better model realistic system behaviors, possibly including tasks involving thermal states and open system dynamics.

In the next Section we discuss about some useful representations for parameterizing quantum channels 
and introduce our parameter optimization protocol.
Then, in Section~\ref{sec:numres} we show some numerical results about the application of the protocol 
to a simple toy case of $\textrm{CNOT}$ optimization with asymmetric noise between the qubits,
while we outline some future perspective in Section~\ref{sec:conclusions}.

\section{Parametric quantum channels: the framework}\label{sec:framework}
While quantum circuits, representing ideally unitary operations mapping 
input pure states into output pure states, can be easily parameterized by the angles of rotation gates, the construction of parametric quantum channels might involve 
also non-unitary processes such as measurements, partial traces and the use of 
stochastic ensambles of other sub-channels or sub-circuits. 
The only assumption we enforce on the protocols implementing a quantum channel 
is that it models a completely positive and trace preserving (CPTP) 
linear map between density matrices~\cite{Nielsen:2012yss}.
Non-trace-preserving channels can in principle be realized through post-selection by discarding 
the state when some specific measurement occur, 
but this is beyond the scope of this proceeding, which focuses instead on quantum channels
where output states are linearly related to input states\footnote{While the output of a 
quantum operation with post-selection can be always normalized back to unit trace, the dependence of
the on the input density matrix is effectively non-linear, 
so that it cannot be described with the formalism of linear superoperators.}.
In the following section we briefly describe 
some useful formal representations and, more importantly, practical realizations of parametric
quantum channels.
It is important to stress that, even in absence of quantum noise from actual hardware,
realizing quantum channels can be useful to gain more control on hybrid quantum-classical approaches,
so here we consider their non-unitarity a resource for computation (common perspective, for example,
in the context of reservoir computing~\cite{PhysRevE.107.035306} or thermal state preparation~\cite{Clemente:2024sft}), not as a drawback which brings a loss in controllability.

\subsection{Quantum channel representations}\label{subsec:qchan-reps}
As mentioned above, even in the absence of noise from the hardware, 
one can still realize CPTP linear maps by engineering the source of 
non-unitarity in different ways, some of which are described in the next sections. 
However, this deviation from unitarity should not be considered as a form of noise, 
since it is actually fully controllable by the specific protocol and parameterization 
of the quantum channel.
On the other hand, the noise coming from running on actual hardware is not directly controllable
and tends in general to distort the engineered channel.
To distinguish between these two situations, 
while we denote by $\mathcal{E}$ an ideal noiseless quantum channel, 
we use $\tilde{\mathcal{E}}$ to indicate the actual noisy quantum channel realized by the hardware
\footnote{Here we make the strong assumption that possible time-dependent variations in 
the noisy quantum channel are negligible compared to deviations from the corresponding ideal noiseless channel.}. In general, even two circuit implementations which would ideally correspond to the same
ideally unitary channel might differ when executed on the hardware.

\paragraph{Kraus representation}
A common representation for a quantum channel $\mathcal{E}$ acting on states in a Hilbert space 
$\mathcal{H}$ with dimension $d\equiv 2^q$ is given by the 
\emph{Kraus decomposition}~\cite{Nielsen:2012yss,Biamonte:2019uzx} in terms of a set of operators $\{K_\alpha\}_{\alpha=0}^{d^2-1}$:
\begin{equation}
    \rho' = \mathcal{E}(\rho) = \sum_{\alpha} K_{\alpha} \rho K_{\alpha}^{\dagger},
\end{equation}
where the trace-preserving property is satisfied if $\sum_\alpha K_\alpha^\dagger K_\alpha =\mathds{1}$.
In general, any unitarily transformed set $K'_{\alpha}=\mathscr{U}_{\alpha}^{\beta} K_{\beta}$, with $\mathscr{U}\in SU(d^2)$, represents the same quantum channel, but it is possible to select a single representative of the equivalence class by choosing a set that satisfies the canonical orthogonality relation $\Tr(K_\alpha^\dagger K_\beta) = \lambda_\alpha \delta_{\alpha \beta}$, where $\lambda_{\alpha}$ are non-increasing~\cite{Bengtsson:2006,Biamonte:2019uzx}.

\paragraph{Stinespring representations}
While the Kraus representation can be useful for quantum information purposes 
in performing formal computations, a more practical approach to quantum channel engineering 
can be realized by performing unitary operations on an extension of the the original Hilbert space 
via a tensor product with an auxiliary space $\mathcal{A} \otimes \mathcal{H}$, 
with measurements only on the space $\mathcal{H}$.
This realization corresponds to the \emph{Stinespring representation} of the quantum channel~\cite{Nielsen:2012yss,Biamonte:2019uzx}, 
which is formally written as 
\begin{equation}
    \rho' = \mathcal{E}(\rho) = \text{Tr}_{\mathcal{A}}\Big[U\Big((\ketbra{\nu_0})_{\mathcal{A}} \otimes \rho\Big) U^\dagger\Big],
\end{equation}
where $\ket{\nu_0}$ can be any state (usually $\ket{0}$ is chosen), 
while $U\in SU(2^{q_s+q_a})$, with $q_s$ and $q_a$ denoting the system and ancilla qubits.
The most general channel could then be realized using $q_a=2 q_s$ qubits, 
therefore requiring thrice the number of qubits with respect to the original ones. 
However, since parameterizing a subset of quantum channels might be sufficient for one's purposes, 
it is useful to consider a smaller number of ancilla qubits $q_a \ll q_s$.
In this representation, non-unitarity comes from the entanglement possibly generated by $U$ between the two subsystems $\mathcal{A}$ and $\mathcal{H}$, which produces an output mixed state when $\mathcal{A}$ is traced out.
Regarding parameterization, it can be realized as customary for unitary circuits in terms of angle 
parameters of rotation gates appearing in $U$.
In terms of the Kraus representation, $U$ and $\{K_\alpha\}$ are related by
\begin{align}\label{eq:Stine-Kraus_relation}
    U \ket{\nu_0}\otimes \ket{\psi} = \sum_{\alpha=0}^{d^2-1} \ket{\alpha}\otimes K_\alpha \ket{\psi} \quad \forall \ket{\psi}.
\end{align}

\paragraph{Stochastic representations}
A class of quantum channels, named \emph{mixed-unitary channels}~\cite{Audenaert::2008}, 
can be realized as an average of the application of an ensemble of unitary circuits $U$ 
distributed according to some probability weight $w(U)\geq 0$, namely
\begin{equation}
    \rho' = \mathcal{E}(\rho) = \bigintsss_{\mathrm{SU}(2^q)}\!\!\!\!\!\! [w(U) dU]\;  U \rho U^{\dagger} \simeq \frac{1}{M} \sum_{\substack{i=1\\ U_i\sim w}}^M U_i \rho U_i^{\dagger},
\end{equation}
where the last term is an approximation of the mixed-unitary channel using a finite number
of $M$ circuits $\{U_i\}$ identically and independently sampled with probability distribution $w(U)$.
In this setup, the parameterization of the quantum channel enters through the probability distribution $w_{\theta}(U)$, which can be chosen to have support only on a specific subset of $\mathrm{SU}(2^q)$, as we show in an example of application in Section~\ref{sec:numres}.
In general, the mixed-unitary class of quantum channels does not parameterize the whole space of CPTP maps.  However, stochastic protocols are particularly useful for their simplicity and can be combined 
with partial Stinespring realizations (i.e., with $q_a<2 q_s$, as defined above), which allow them to reach beyond the mixed-unitary class.

\subsection{Parameter optimization}
In the following discussion, we write any parameter 
either appearing as an angle in rotation gate in the case of Stinespring realizations
or as parameterizing the probability distribution of quantum circuit 
ensambles in the stochastic realization, with a symbol $\theta_i$, while $\Theta$ indicates 
the full set of parameters.

\noindent
To approximate a target unitary channel $U \in \mathrm{SU}(2^q)$, 
we construct a parametric quantum channel $\mathcal{E}^{(\Theta)}$. The optimization objective minimizes the distance between the ideal ($\mathcal{E}_U$) and realized ($\tilde{\mathcal{E}}^{(\Theta)}$) channels in the form of a cost function
\begin{equation}\label{eq:cost-func}
    C(\Theta) = d(\mathcal{E}_U, \tilde{\mathcal{E}}^{(\Theta)}),
\end{equation}
for some definition of distance between quantum channels.
An useful metric for comparing two quantum channels $\mathcal{E}_1$ and $\mathcal{E}_2$ 
is the \emph{diamond norm}~\cite{Aharonov::1998,Kitaev::2002,Gilchrist::2005,Benenti::2010,Watrous::2018}, expressed as
\begin{equation}\label{eq:diamond-dist}
    d_{\diamond}(\mathcal{E}_1, \mathcal{E}_2) = \sup_{n\leq d, \eta \geq 0} d_{\Tr}\big((\mathcal{I}_n \otimes \Delta\mathcal{E}_{12})(\eta),0\big),\qquad d_{\Tr}\big(\rho_1,\rho_2\big)\equiv \frac{1}{2} \Tr |\rho_1-\rho_2|,
\end{equation}
where $d_{\Tr}$ is the \emph{trace distance} between density matrices~\cite{Nielsen:2012yss}, while $\Delta\mathcal{E}_{12}\equiv \mathcal{E}_1-\mathcal{E}_2$ and $\mathcal{I}_n$ is a trivial extension 
of the $\mathcal{E}_i$ channels on an auxiliary space $\mathcal{A}_n$ of dimension 
$\dim \mathcal{A}_n \equiv n\leq d$ (this space is not the same considered for the Stinespring dilation).
For later convenience, we define the $n$-dependent partial diamond distance before the $\sup_n$ as
\begin{align}\label{eq:n-th-trdist}
    d^{(n)}(\mathcal{E}_1,\mathcal{E}_2) = \sup_{\eta\geq 0} d_{\Tr}\big((\mathcal{I}_n \otimes \mathcal{E}_{1})(\eta),(\mathcal{I}_n \otimes \mathcal{E}_{2})(\eta)\big).
\end{align}
The input density matrix $\eta$ (trace $1$ understood), in common to both channels before computing 
$d^{(n)}(\mathcal{E}_1,\mathcal{E}_2)$, is a generally mixed state in the space 
$\mathcal{A}_n\otimes \mathcal{H}$, while the output density matrices between 
the $\mathcal{I}_n\otimes \mathcal{E}_i$ channels are compared through the trace distance $d_{\Tr}$.
As discussed in~\cite{Kitaev::2002,Gilchrist::2005}, considering general trivial extensions, instead of
directly evaluating the $\sup$ over states on $\mathcal{H}$, is essential since the input state 
fed to a quantum channel $\mathcal{E}_i$ might be in generic entanglement with the environment.
While increasing the dimension of the extension $n$ allows for more and more entangled states,
saturating to maximal entanglement when $n=d$, the behavior of 
$d^{(n)}(\mathcal{E}_1,\mathcal{E}_2)$ of the trace distance as a function of $n$ is not monotonic in general. Therefore, one should in principle compute $d^{(n)}$ for all $n\in [0,d]$, evaluating the 
occurrence of the maximum when the worst $n$ is realized.
Regarding the channel optimization, considering the cost function to be minimized in Eq.~\eqref{eq:cost-func}, for each $n$ the task consists in finding the optimal $\Theta^*$ which saturate $\min_{\Theta}\max_{n} d^{(n)}(\mathcal{E}_U,\tilde{\mathcal{E}}^{(\Theta)})$.
In practice, by the double convexity of the distances, it derives~\cite{Gilchrist::2005} 
that for any $n$ the $\sup_{\eta\geq 0}$ is saturated by $\eta$ being pure states 
(since these states are at the boundary of the density matrices, with mixed states at the bulk),
so they can be prepared for example by parameterizing a general unitary circuit $V(\alpha)$ 
and minmaxing the quantity 
\begin{align}\label{eq:minmax}
    C^{(n)}(\Theta;\alpha) \equiv d_{\Tr}\big((\mathcal{I}_n \otimes \mathcal{E}_{U})(\eta_\alpha),(\mathcal{I}_n \otimes \tilde{\mathcal{E}}^{(\Theta)})(\eta_\alpha)\big), \quad\text{with}\quad \eta_\alpha\equiv V(\alpha)\ketbra{0}V^\dagger(\alpha).
\end{align}
This optimization can go through a jointed (possibly approximate) gradient descent-ascent for $\Theta$-$\alpha$ to obtain $\min_{\Theta} \max_{\alpha} C^{(n)}(\Theta;\alpha)$.
Notice that evaluating the trace distance in Eq.~\eqref{eq:diamond-dist} or Eq.~\eqref{eq:minmax} 
on a real quantum hardware requires some form of process or state tomography and can be done
only with a finite number of shots. Therefore, an approximate approach such as \emph{classical shadows} 
estimate of mixed states properties would be preferable on larger systems~\cite{Huang:2020tih}.

\section{A himple example of application: stochastic $\textrm{CNOT}$ with asymmetric noise}\label{sec:numres}

\begin{figure}[h!]
    \centering
    \includegraphics[width=0.8\textwidth]{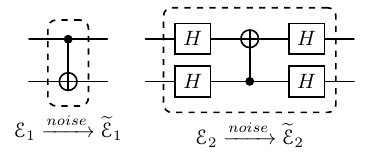}
    \caption{Two possible, ideally equivalent, realizations of a $\textrm{CNOT}$ unitary between qubit $0$ (top) and qubit $1$ (bottom).}
    \label{fig:cnot_both}
\end{figure}

Here we show a mixed-unitary channel realization and optimization where the target unitary 
is a CNOT gate between qubit $0$ and qubit $1$ (i.e., $U\equiv C_0 X_1$ $\implies$ $\mathcal{E}_{\textrm{CNOT}}(\rho)\equiv (C_0 X_1) \rho (C_0 X_1)$) in the presence of asymmetric noise between the two qubits involved.
The mixed-unitary channel we consider is built as a weighted mixture from only two very 
simple implementations, shown in Figure~\ref{fig:cnot_both}, 
of what would be the same ideally unitary target $\textrm{CNOT}$.
In practice, the effect of noise on single qubit gates is often negligible, so, in the presence of symmetric noise (i.e., if the two qubits are affected by similar noise contributions),
one does not observe but marginal advantages in using one or the other variant.
With asymmetric noise, the effective channels $\tilde{\mathcal{E}}_1$ 
and $\tilde{\mathcal{E}}_2$ differ, so one can investigate which convex combination
\begin{equation}
    \tilde{\mathcal{E}}^{(w_1)} = w_1 \tilde{\mathcal{E}}_1 + (1 - w_1) \tilde{\mathcal{E}}_2
\end{equation}
minimizes the cost in Eq.~\eqref{eq:cost-func} as a function of the only channel parameter $w_1\in [0,1]$, according to the stochastic channel representation introduced in Sec.~\ref{subsec:qchan-reps}.
The platform we considered for the runs is the \emph{Pennylane} library with \texttt{default.mixed} emulator~\cite{Bergholm:2018cyq}, where we added noise sources using \texttt{qml.DepolarizingChannel} and \texttt{qml.AmplitudeDamping} on both qubits individually before and after the two $\textrm{CNOT}$s in both versions\footnote{In this simple test we are ignoring other possible multi-qubit noise channels, which would likely contribute with additional asymmetric sources.}, 
where coefficients have been arbitrarily chosen according to Table~\ref{tab:noise_params}
(other asymmetric choices exhibit the similar phenomenology, but the gain from the mixed strategy
starts to disappear when getting closer to the symmetric case). 
\begin{table}[h]
    \centering
    \begin{tabular}{lcc}
        \toprule
        \textbf{Noise Type} & \textbf{Qubit 0 coefficient} & \textbf{Qubit 1 coefficient} \\
        \midrule
        Depolarizing &  0.01  & 0.03 \\
        Amplitude Damping & 0.05 & 0.3 \\
        \bottomrule
    \end{tabular}
    \caption{Noise parameters for single-qubit noise channels chosen asymmetrically for the two qubits. The coefficients values are related to the single parameters $p$ and $\gamma$ appearing in the definition of \texttt{qml.DepolarizingChannel} and \texttt{qml.AmplitudeDamping} respectively.}
    \label{tab:noise_params}
\end{table}
\begin{figure}[h!]
  \centering
  \includegraphics[width=1.0\textwidth]{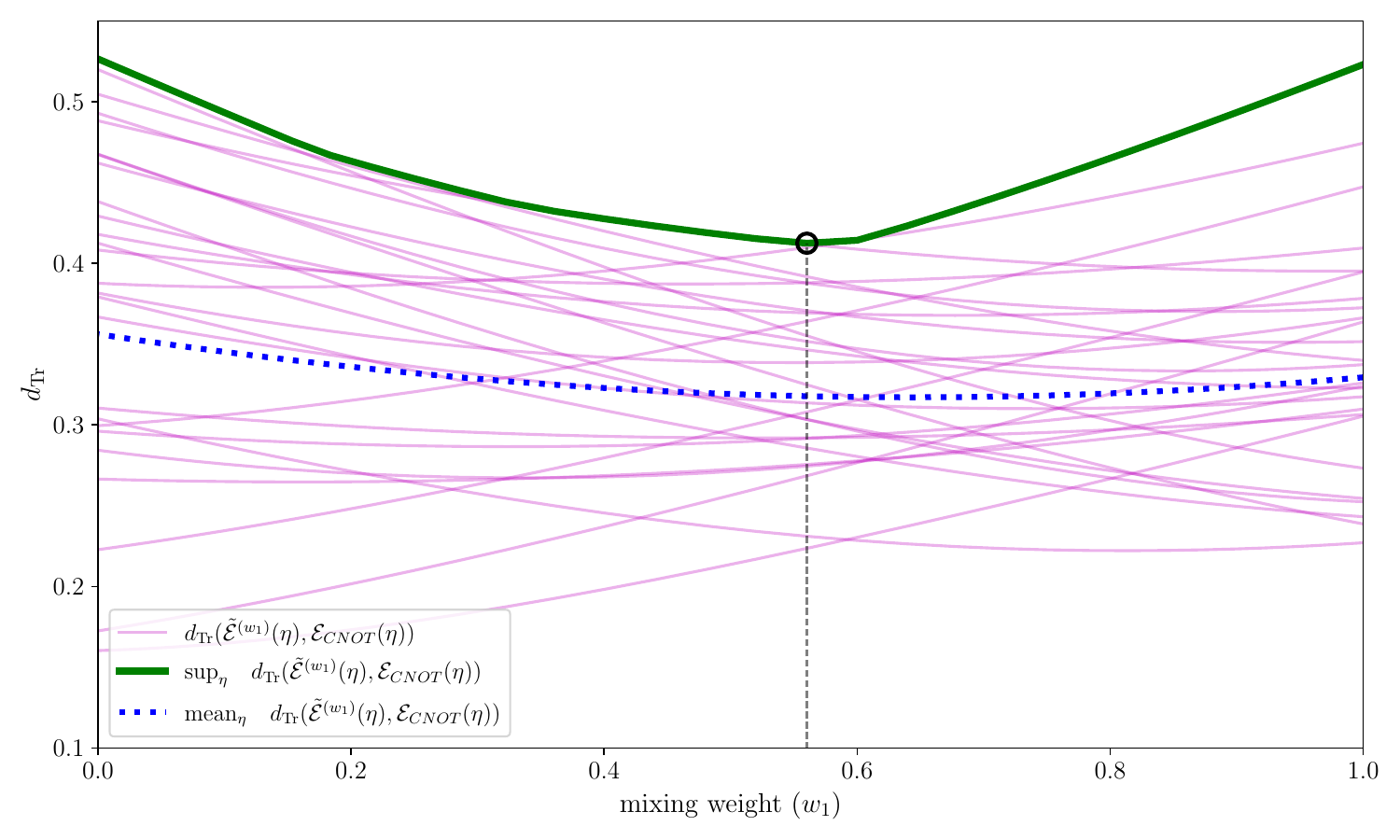}
  \caption{Cost function $C^{(0)}(w_1;\alpha)$ as a function of $w_1$. Thin magenta lines correspond to some different initial input states.  The behavior for the $\textrm{sup}_\eta$ and the $\textrm{mean}_\eta$ at any fixed value of the mixing parameter $w_1$ is represented respectively by 
  the thick green line and the blue dotted line, while the empty black dot corresponds to the min-max solution, which is reached through minmaxing via gradient descent-ascent.}
  \label{fig:cost_plot}
\end{figure}
The behavior of the cost function $C^{(0)}(w_1;\eta)$ is shown in Figure~\ref{fig:cost_plot}. 
This contains both the worst case for different values of $w_1$ (thick green line), where we can find 
the optimal $w_1$ as its minimum, as well as (thin magenta lines) the general behavior of the mixed channel for a few (pure) input states $\eta=\ketbra{\psi_0}$ where $\ket{\psi_0} = V\ket{0}$ and $V$ is sampled randomly from $\mathrm{SU}(4)$ with the Haar measure. Even if not necessary for the actual minmax optimization, this illustrates quite nicely that the single circuits at the border of the mixing domain (i.e., $w_1=0$ or $w_1=1$) can behave better than for the ``optimal'' $w_1^*$ solution for at least some input
states, but since this channel is supposed to take part of a generic stage of a much larger circuit, 
one cannot really control which input is fed, making the minmax solution the more robust and consistent one, even if not necessarily the best one for certain runs. The average distance instead appears to exhibit a quite milder behavior, where the ``minmean'' solution for $w_1$ seems also sufficiently close to the 
minmax one.
Moreover, at least for this simple example, we tested that for trivial extensions ($n>0$) the costs $C^{(n)}$ are not higher than $C^{(0)}$, so the worst case for $C=\sup_n C^{(n)}$ is already realized by $C^{(0)}$.

\section{Conclusions and outlook}\label{sec:conclusions}
Parameterized quantum channels extend unitary computing to NISQ devices by adapting to noise properties through parametric optimization. They generalize standard error mitigation techniques and can be tailored for specific tasks where a quantum channel has to be reused multiple times as component of a longer circuit. This includes, for example, real-time evolution with Trotterization, where the same basic block of unitary circuit is repeated at many locations of the full circuit. Future directions include also addressing trainability and scaling on real hardware, investigating the effect of crosstalks between different qubits, as well as considering generalizations of the parametric optimization in the case of non-linear quantum operations (e.g., via postselection).

\acknowledgments
GC acknowledges support from the National Centre on HPC, Big Data and Quantum Computing - SPOKE 10 (Quantum Computing) and received funding from the European Union Next-GenerationEU - National Recovery and Resilience Plan (NRRP) – MISSION 4 COMPONENT 2, INVESTMENT N. 1.4 – CUP N. I53C22000690001.

\end{document}